\documentclass[prc,aps,showpacs]{revtex4}
\usepackage{graphicx}
\begin{document}

\title{Constraints on narrow exotic states
from $K^+p$ and $K^0_L p$ scattering data}
\author{R.~L.~Workman}
\email{rworkman@gwu.edu}
\author{R.~A.~Arndt}
\author{I.~I.~Strakovsky}
\affiliation{Department of Physics, The George Washington 
             University, Washington, D.C. 20052-0001}
\author{D.~M.~Manley}
\email{manley@kent.edu}
\author{J. Tulpan}
\affiliation{Department of Physics, Kent State University, 
Kent, OH 44242-0001} 

\begin{abstract}

We consider the effect of exotic S=+1 resonances 
$\Theta^+$ and $\Theta^{++}$ on $K^+p$ elastic scattering 
data (total cross section) and the process $K^0_Lp\to K^0_Sp$. 
Data near the observed $\Theta^+(1540)$ are examined for 
evidence of additional states.  The width limit for a 
$\Theta^{++}$ state is reconsidered. 

\end{abstract}


\maketitle

The observation of a narrow exotic S=+1 resonance, the $\Theta^+(1540)$,
has now been documented in the most recent edition of the Review of Particle
Physics~\cite{PDG}. 
Its initial discovery at SPring-8 and subsequent apparent confirmation
at facilities worldwide~\cite{spr8,itep,jlab,barth}
required the re-establishment of a PDG category
(the $Z^\ast$ states) long since removed due to a lack of new measurements
and insufficient theoretical justification. This change in attitude was
largely due to the verification of a predicted narrow state. The broad
Z$^\ast$ states observed earlier were at least qualitatively described by
a non-exotic coupling to $K \Delta$ and $K^\ast N$ states.

Given the implications of this discovery for baryon spectroscopy, 
a massive theoretical and experimental effort has been launched to 
verify the result and to search for associated states. Conclusions from
this second wave of activity have been mixed at best. High-energy
facilities have produced a number of negative results and the model
basis for the initial prediction~\cite{dpp} has been criticized~\cite{cohen}. 
These developments
have led to repeated studies with higher statistics and a more careful
handling of systematic effects. 

The most direct method to produce a resonance decaying to $KN$ would
be through kaon-nucleon and kaon-nucleus scattering. The ITEP result,
scattering in a Xe bubble chamber, produced a tight width limit~\cite{itep}
associated with the observed peak ($\Gamma$ $<$ 9~MeV). 
The bulk of existing
kaon-deuteron scattering and breakup data show little evidence for a
resonance structure, leading to width limits below 
the 1 -- 2~MeV level~\cite{CaT,nus,GW1,Haid1,Haid2}. 

A reanalysis of the ITEP experiment~\cite{itep}
by Cahn and Trilling~\cite{CaT} has resulted in a finite width
of approximately 1 MeV, with a similar value coming from
Gibbs who has interpreted a bump in the $K^+ d$ total 
cross section data as evidence for a narrow resonance~\cite{Gibbs}.
While these results have been challenged~\cite{Haid2}, we take
the above estimates as a guide in searching for other associated
states.

Searching for a $\Theta^{++}$ signal in the existing $K^+p$ scattering data
is much simpler task. 
Over a wide energy range (1500~MeV to 1700~MeV), associated
with the $\Theta^+$ and a possible related $\Theta^{++}$, 
the $K^+p$ total cross
section is essentially flat and is covered by a number of independent
(and consistent) measurements~\cite{kpref}. 
In a previous search~\cite{GW1}, we scanned the database
by inserting narrow states in a number of partial waves, and over a range
of energies, to search for an improved fit (discounting the datapoints
near to an added peak). In this study, we claimed to see no evidence for
a $\Theta^{++}$ state, but provided a width limit only for the 
$\Theta^+(1540)$.  

\begin{figure}[th]
\centering{
\includegraphics[height=0.7\textwidth, angle=90]{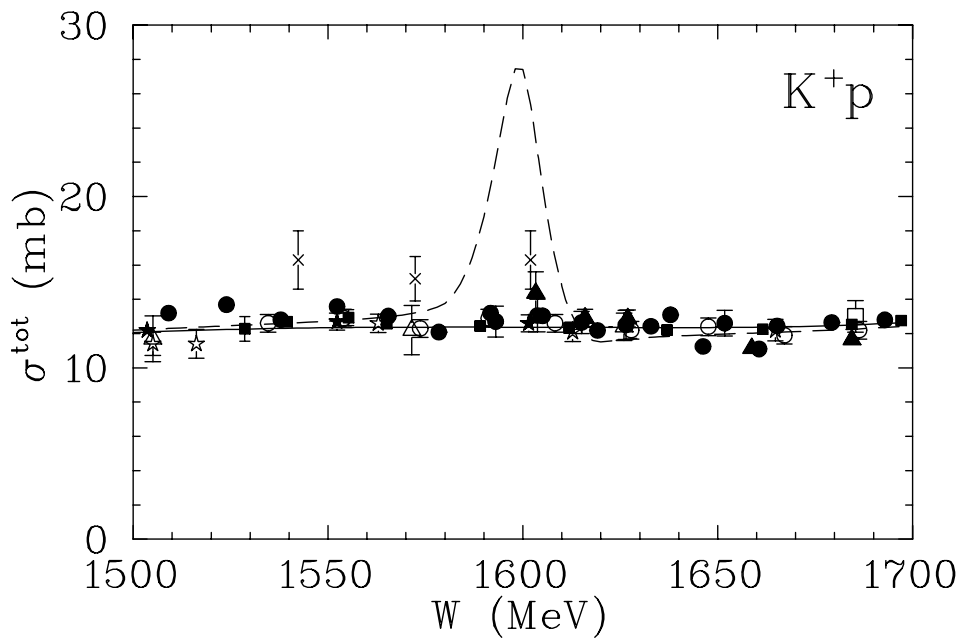} 
}\caption{Total cross section for $K^+p$ scattering 
          from Ref.~\protect\cite{kpref} compared to the 
          fit of Ref.~\protect\cite{Z0} with (dashed) and 
          without (solid) an elastic $\Theta^{++}$ resonance 
          in the $P_{13}$ partial wave,
          having a 1.6~GeV mass and 1~MeV width.  Curves 
          account for beam-momentum resolution. Data
          statistical and systematic uncertainties have been 
          added in quadrature. \label{fig:g1}}
\end{figure}

A more appropriate method, in the case of $K^+p$ scattering,
would be to add a resonance peak taking into account the momentum spread
of the incident kaon beam. This effect is usually ignored when searching for
broad states, but is crucial in cases when the resonance width is 
comparable to (or less than) the momentum resolution of the incident beam.
A representative example is given in 
Fig.~\ref{fig:g1} for a beam with a momentum spread of 
30~MeV/$c$ (FWHM)~\cite{kpref} producing a $P_{13}$ resonance~\cite{jvalue}
at 1600~MeV with a width of 1~MeV. Here the resonance has been directly
added to the fit of Ref.~\cite{Z0} to give an order-of-magnitude result.
We have also assumed the $\Theta^{++}$ to be 
an elastic resonance; no estimate is possible for
a state decaying dominantly into another channel. With these assumptions,
the 1 -- 2~MeV limit associated with the $\Theta^+$
is clearly much too large for a $\Theta^{++}$ candidate. 
If such a state exists, and decays mainly into
$K^+p$, its width would necessarily be significantly less than 1~MeV. 

In Ref.~\cite{man1}, we applied this method to a study of the $K^0_L p$
total cross section. The amplitude for this process is given by
$M_{K_L^0p}\; = \; (Z_0 + Z_1 + 2 Y_1)/4$, where $Z_{0,1}$ are the 
strangeness S = 1, I = 0 and 1 amplitudes, and $Y_1$ is the S = $-1$, I = 1
amplitude. While sensitive to both isoscalar and isovector pentaquark
candidates, the use of $K_L^0$ scattering requires knowledge 
of a sizable isovector
S = $-1$ amplitude. This uncertainty and the rather large statistical errors
associated with the existing 
data combined to prevent any conclusions about the exotic resonance content.

A more promising reaction is $K^0_L p \to K^0_S p$. The available integrated
cross sections for this process have much smaller statistical errors than
those associated with
the $K^0_L p$ total cross sections. The existence of angular distribution
data is another advantage, as they could reveal the angular momentum 
associated with a $\Theta$ resonance decay. The amplitude is
given by $M_{K_L^0 p \to K_S^0 p}\; = \; (Z_0 + Z_1 - 2 Y_1)/4$ and is
therefore also sensitive to both S = +1 and S = $-1$ resonances. 

\begin{figure}[th]
\centering{
\includegraphics[height=0.7\textwidth, angle=90]{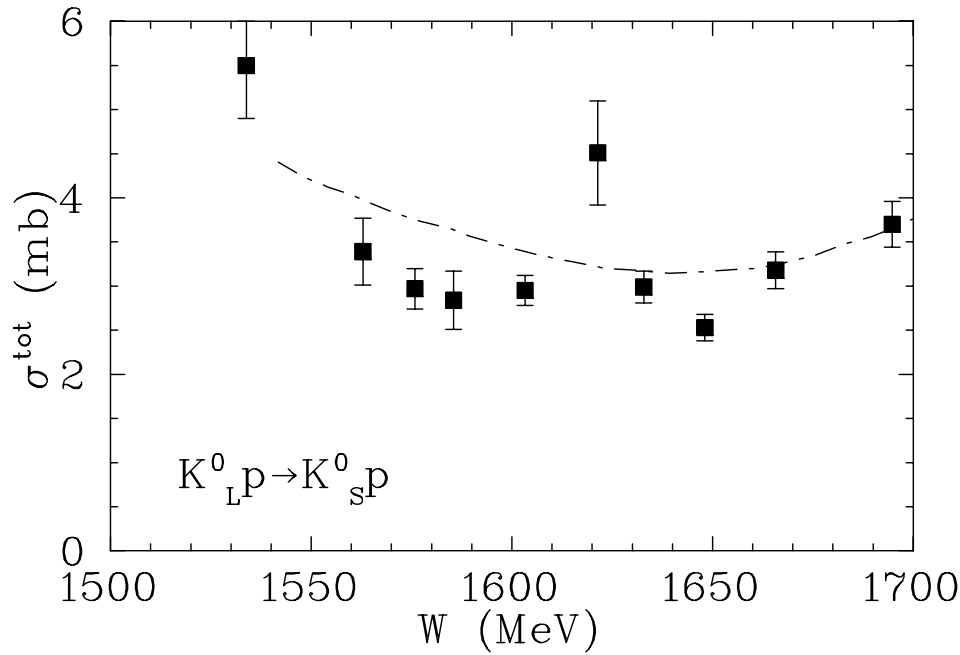}
}\caption{Integrated cross section for $K^0_Lp\to K^0_Sp$
          from Ref.~\cite{bigi} compared to a fit, given in
          Ref.~\cite{bigi}, based on the amplitudes of
          Refs.~\cite{Y1,martin}. \label{fig:g2}}
\end{figure}

\begin{figure}[th]
\centering{
\includegraphics[height=0.7\textwidth, angle=90]{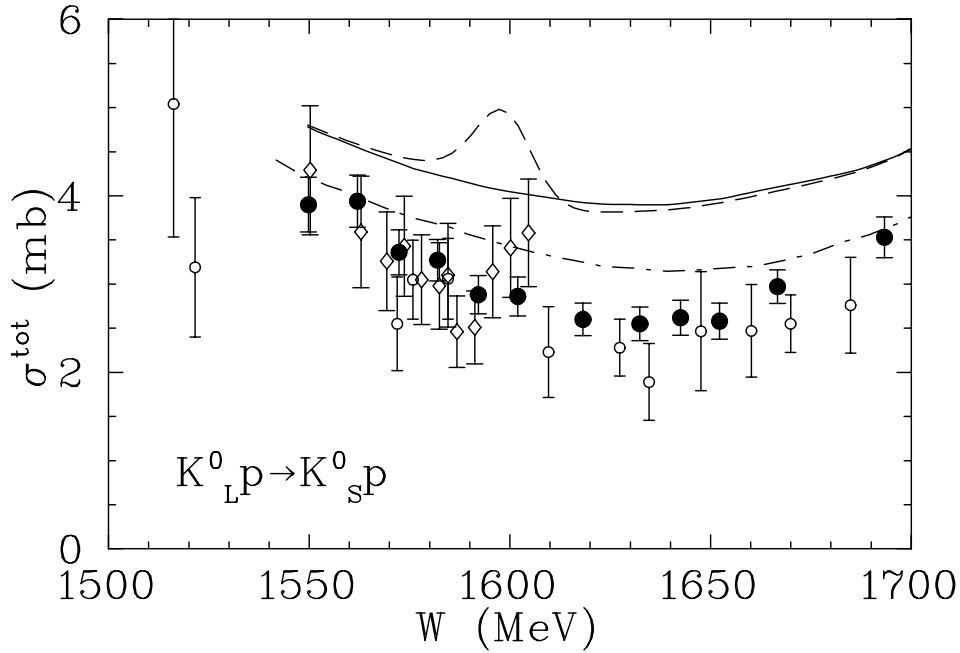}
}\caption{Integrated cross section for $K^0_Lp\to K^0_Sp$ 
          from Ref.~\protect\cite{cameron,klsref} compared 
          to a prediction based on the amplitudes of 
          Refs.~\protect\cite{Z0,Y1} with (dashed) and 
          without (solid) an added $\Theta^+$ resonance at 
          1.6~GeV with a width of 2~MeV.  The dot-dashed 
          curve~\protect\cite{cameron} is based on the 
          amplitudes of Ref.~\protect\cite{Y1,martin}, allowing 
          the $Z_0$ amplitudes (S-wave) to vary. Data statistical
          and systematic uncertainties have been added in
          quadrature.  \label{fig:g3}}
\end{figure}

Our examination of this reaction was prompted by the measurement of 
Ref.~\cite{bigi}, which appears to have a sharp structure slightly above
1600 MeV. This data is plotted in Fig.~\ref{fig:g2}. Unfortunately, 
the most precise measurements of Ref.~\cite{cameron}, which begin at 
the $\Theta^+$ mass, show no sign of this structure. A less pronounced
bump could be supported by the data of Ref.~\cite{klsref}.
The energy dependence from 1500~MeV to 1700~MeV is shown in 
Fig.~\ref{fig:g3}.

A direct computation of the
integrated cross section, using the amplitudes of Refs.~\cite{Z0,Y1}
reproduces the overall shape but not the normalization. This effect
was also noted in Ref.~\cite{bigi,cameron}. Using the amplitudes of 
Refs.~\cite{Y1,martin} some improvement was found in Ref.~\cite{bigi}
by allowing the S-wave $Z_0$ amplitudes to vary and become inelastic.
As this solution is likely to violate unitarity, particularly at low
energies, we have treated the normalization as an unknown, and have
concentrated on the effect of resonances in the higher partial waves.

The expected size of a $\Theta^+$ resonance contribution is given in
Fig.~\ref{fig:g3}, assuming a beam-momentum distribution with FWHM of 
30~MeV/$c$~\cite{cameron}.
The resonance has been placed arbitrarily at 1600 MeV, in the $P_{01}$
$KN$ partial-wave, assuming a width of 2~MeV and a decay into only $KN$
final states. This has been superimposed on the smoothly varying 
result calculated using the amplitudes (unmodified) from Refs.~\cite{Z0,Y1}.
The absence of a structure of this magnitude in the data should not be
taken as conclusive evidence against a resonance. One should note that there
exists a 4-star $\Sigma (1670)$ resonance that is similarly invisible in
Fig.~3. 

In summary, we have extended our investigations of two-body reactions which
could potentially give evidence for or against an exotic $\Theta$ resonance.
No conclusive result was found. Some evidence for a structure around 1600 MeV
appears in the integrated cross section for 
$K^0_L p \to K^0_S p$, but is absent in the most precise set of measurements
covering this region. [One should note that a plot~\cite{pdg_plot} 
of the $K^+$-deuteron 
total cross section has a fluctuation at about the same lab beam momentum.]
In the case of a $\Theta^{++}$ 
state, we have quantified the statement given in Ref.~\cite{GW1}.
If the $\Theta^{++}$ exists and has a $K^+ p$ decay width comparable to
the (1 MeV) width associated with the $\Theta^+$, it should have been 
seen in the existing $K^+ p$ scattering data. Here we have assumed $K^+ p$
to be the dominant decay channel, following the suggestion of 
Ref.~\cite{capstick}.

\acknowledgments

This work was supported in part by the U.~S. Department 
of Energy Grants DE--FG02--99ER41110 and 
DE--FG02--01ER41194. R.~W. and I.~S. gratefully 
acknowledge a contract from Jefferson Lab under which 
this work was done.  Jefferson Lab is operated by the 
Southeastern Universities Research Association under the 
U.~S.~Department of Energy Contract DE--AC05--84ER40150.


\vfil
\eject


\begin{thebibliography}{99}
\bibitem{jvalue} In this simple approximation, assuming negligible background
               interference, the resonance peak scales as (2J+1). For a state 
               in the $K^+ p$ $S_{11}$ partial wave, however, there would be 
               significant interference.
\end{thebibliography}
\end{document}